\documentclass[conference]{IEEEtran}
\IEEEoverridecommandlockouts
\usepackage{cite}
\usepackage{amsmath,amssymb,amsfonts}
\usepackage{graphicx}
\usepackage{bbding}
\usepackage{textcomp}
\usepackage{xcolor}
\usepackage{algorithm}
\usepackage{amsmath}
\usepackage[noend]{algpseudocode}
\usepackage{hyperref}
\usepackage{booktabs}
\usepackage{stfloats} 
\hypersetup{
    colorlinks=true,
    linkcolor=black,
    citecolor=black,
    urlcolor=blue,
    pdfborder={0 0 0}
}

\def\BibTeX{{\rm B\kern-.05em{\sc i\kern-.025em b}\kern-.08em
    T\kern-.1667em\lower.7ex\hbox{E}\kern-.125emX}}
\begin{document}

\title{ i-LAVA: Insights on Low Latency Voice-2-Voice Architecture for Agents\\
}

 \author{\IEEEauthorblockN{Aditya Choudhary \IEEEauthorrefmark{2} }
 \IEEEauthorblockA{ \textit{Sprinklr AI } \\
 Gurugram, India \\
 }

\and
\IEEEauthorblockN{Anupam Purwar\IEEEauthorrefmark{1}\IEEEauthorrefmark{2}}
\IEEEauthorblockA{\textit{Sprinklr AI} \\
Gurugram, India \\
}\thanks{\IEEEauthorrefmark{1}Corresponding Author: Anupam Purwar (e-mail: anupam.aiml@gmail.com, https://anupam-purwar.github.io/page/)}
\thanks{\IEEEauthorrefmark{2} Equal contributions by both authors}

}

\maketitle

\begin{abstract}
We experiment with a low-latency, end-to-end voice-to-voice communication model to optimize it for real-time conversational applications. By analyzing components essential to voice to voice (V-2-V) system viz. automatic speech recognition (ASR), text-to-speech (TTS), and dialog management, our work analyzes how to reduce processing time while maintaining high-quality interactions to identify the levers for optimizing V-2-V system. Our work identifies that TTS component which generates life-like voice, full of emotions including natural pauses and exclamations has highest impact on Real time factor (RTF). The experimented V-2-V architecture utilizes CSM1b has the capability to understand tone as well as context of conversation by ingesting both audio and text of prior exchanges to generate contextually accurate speech. We explored optimization of Residual Vector Quantization (RVQ) iterations by the TTS decoder which come at a cost of decrease in the quality of voice generated. Our experimental evaluations also demonstrate that for V-2-V implementations based on CSM most important optimizations can be brought by reducing the number of RVQ Iterations along with the codebooks used in Mimi.

\end{abstract}


\section{Introduction}
With the advent of LLMs and their adoption for business use-cases, several text based Customer Support Channels and other interaction domains have moved to extensive use of AI Agents \cite{bandraupalli2025vlmsinthewildbridginggapacademic} \cite{verma2025mpacemotherchildframework}. Retreival augmented generation based agents powered by  Language models have already demonstrated huge promise in knowledge based agents for question answering tasks and \cite{juvekar2024introducingnewhyperparameterrag} \cite{purwar2023keywordaugmentedretrievalnovel}. Besides, the ability of LLMs to natively ingest and output text data as well as multimodal data motivates the widespread use of agents by ensuring low latency responses powered by the reasoning capabilities \cite{b2024evaluatingefficacyopensourcellms} \cite{gupta2025earmoredgecaseassessment}. The next frontier for adoption of AI Agents are the channels requiring voice data as input and output. The present state of Multi-Modal LMs lacks the maturity of LLMs in terms of their reasoning and tool-calling capabilities, both essential for an AI Agent \cite{sriram2025mindthework}. Hence, low latency voice-to-voice models still hold their relevance for industrial use cases viz. customer support voice agents. It is paramount that the Agent is contextually aware to the tone of speech and responds accordingly in a human like voice. We present our work on integrating a Residual Vector Quantization (RVQ) based Text-to-Speech (TTS) model optimized for a low latency pipeline. Our work presents an end-to-end pipeline which takes the above mentioned requirements into account for a low latency voice-to-voice interaction channel.

\section{Methodology}
Our proposed system consists of 3 important blocks: Speech to Text , LLM inference text to speech and Text to speech (TTS). The end-to-end pipeline (\ref{fig:pipeline}) supports live audio transcription followed by LLM augmented response generation and low latency text-to-speech service delivering an interactive real-time experience.
The ASR component is configured to start transcribing the live speech chunk by chunk and eventually produce a coherent output which accounts for removal of pauses and gap words. Subsequently the output is passed to the LLM component along with the prior conversation transcriptions and LLM generated text responses for context aware replies. The text output of the LLM is fed to the TTS component which is highly optimized for low latency, delivering the first audio chunk in as low as 640.9 ms in the GPU compute environment. The proposed system has been evaluated  over both CPU and GPU environments:
\begin{itemize}
    \item CPU Environment: Apple M4 Max
    \item GPU Environment: NVIDIA L4 (24GB RAM)
\end{itemize}
\subsubsection{Automated Speech Recognition}
We utilized OpenAI's whisper \verb|v3-large-turbo|  model \cite{radford2022whisper} for audio transcription. The model is used with the chunked algorithm for faster processing.
\subsubsection{LLM}
For minimal latency, OpenAI's \verb|gpt-4o-mini| was used as the intermediate LLM for a balance between LLM intelligence and latency. We experimented with both streaming the LLM output and one-shot generation. For streaming, we pass the LLM output to TTS every 10 tokens generated.
\subsubsection{Text to Speech}
We utilize the CSM-1B model \cite{csm}. The model uses Residual Vector Quantization which implements multiple stages of quantization to encode audio data efficiently \cite{kumar2023highfidelityaudiocompressionimproved}. 
It starts with a base quantizer and then subsequent stages quantize the residual errors from the previous stage, creating a highly accurate audio representation. The model is also capable of processing given text and audio context for improved conversation skills .
\subsection{Fast life-like speech}
The standard implementation of CSM-1B uses 32 iterations of RVQ computation (32 codebooks) to generate life-like speech. However, this is a major bottleneck to voice generation speed owing to the process being sequential. For the same, we have applied following optimizations:
\begin{itemize}
    \item \textbf{Code to Kernel:} CSM-1B uses a Llama 3.2 1B as backbone and Llama 3.2 100M as decoder. The open source implementation of the code was optimized by generating Kernels via JIT Compilation using \verb|torch.compile| for the backbone, decoder and other functions
    \item \textbf{RVQ Iterations:} In order to reduce the latency, we decrease the number of iterations of RVQ computation, consequently the number of codebooks used in the Mimi Tokenizer used by CSM is also set to the number of RVQ iterations. We also experiment with reduced number of RVQ iterations and keeping the number of Mimi codebooks at 32 by padding the RVQ output by two of the following methods:
    \begin{itemize}
        \item Mean Padding: Pad the RVQ output by padding the remaining dimensions with the mean of the tokens in the output
        \item Concat Padding: Pad the RVQ output by concatenating the same array to the end
    \end{itemize}
    \item \textbf{Cold Start} A cold start before inference greatly helps both the whisper and the CSM models. CSM-1b takes atleast 2 generations before achieving its desired performance. \\
\end{itemize}

\subsubsection{One Shot Generation}
The model is able to achieve a Real Time Factor (RTF) $<1$ (0.383) on GPU processing without reducing RVQ Iterations and RTF of $<1$ on CPU but with reducing the RVQ Iterations to 16. \\

\subsubsection{Streaming}
Streaming improves the latency of the model by bridging the gap between the user ending their speech and getting to hear the beginning of the response. This is facilitated by chunking the audio generation and hence the latency accounts only for the time it takes to generate the first audio chunk.
A seamless experience is guaranteed by the fact that inter-chunk generation time is lesser than the length of the chunks, hence maintaining a non-empty queue for audio. RTF for streaming is naturally higher than One Shot Generation.

\subsection{End-to-End Pipeline}
    \begin{figure*}[htbp]
        \centering
        \includegraphics[width=0.8\textwidth]{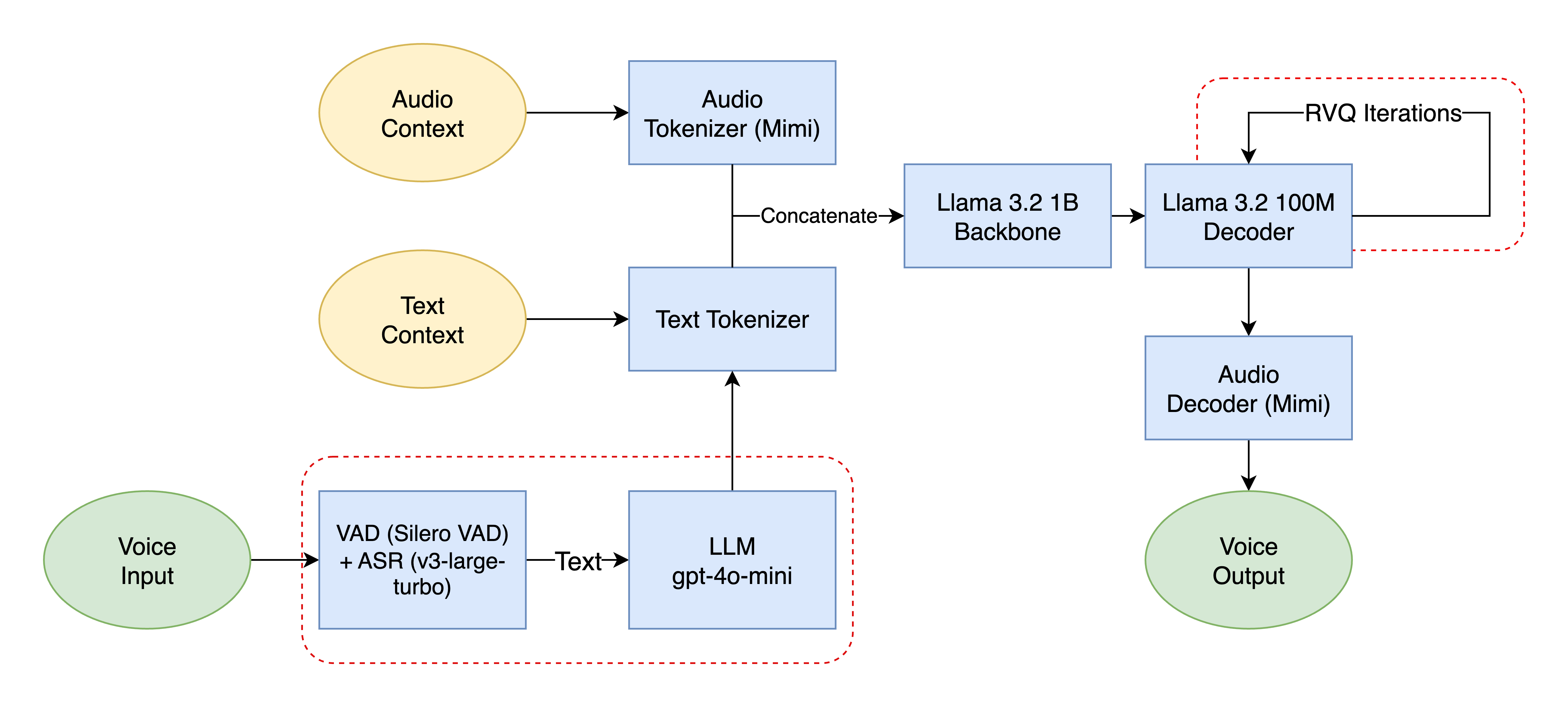}
        \caption{End-to-End Pipeline ingesting the context (Yellow) and the Voice Input comprises of a VAD+ASR component and LLM component (Blue enclosed in Red) for augmenting voice input to transcribed text and subsequently LLM response to be fed into the TTS module (Blue) which processes the tokens through a Llama backbone and decoder architecture with an optimized RVQ Iteration process (Loop enclosed in Red) to eventually yield Voice Output}
        \label{fig:pipeline}
    \end{figure*}
The end-to-end pipeline described in Figure \ref{fig:pipeline} represents the architecture describing the flow of data and the processes involved. A Voice Activity Detector (Silero-VAD) \cite{silero-vad} filters the relevant part of speech and passes it to the ASR component for transcription which is further fed to the LLM for response generation. The context audio, context text and LLM response text are tokenized and concatenated to be fed into the TTS component wherein the tokens are processed through RVQ Iterations. The output of the RVQ process is fed to an Audio Decoder which generates the speech response.

\section{Results}
Our experiments on V-2-V architecture demonstrate that all components VAD, ASR, LLM based inference and finally TTS needs to work in tandem to generate real life like streaming conversation. 

\subsection{LLM Streaming}
We evaluated our proposed architecture end to end by running all 3 components: ASR, LLM inference and TTS.  For the same, we performed time profiling of entire pipleine to measure  the latency contributed by the various components with TTS running on 16 RVQ iterations, the results of time profiling are capture in Table 1.

\begin{table}[htbp]
\caption{Time Profile}
\centering
\begin{tabular}{|l|c|c|c|c|}
\hline
\textbf{Component} & \multicolumn{2}{|c|}{\textbf{GPU}} & \multicolumn{2}{|c|}{\textbf{CPU}} \\
\hline
LLM Streaming & \XSolidBrush & \checkmark & \XSolidBrush &  \checkmark \\
\hline
ASR (s) & 0.5056 & 0.4986 & 2.4502 7 & 2.2245\\
LLM (s) & 2.1502 & 0.1504 & 1.3272 & 0.1496 \\
TTS (s) & 0.6695 & 0.6345 & 1.6299 & 1.4647 \\
Total (s) & 3.3253 & 1.2835 & 5.4073 & 3.8388 \\
\hline
\end{tabular}
\label{tab:time_profile}
\end{table}


Note that the data corresponds to a single instance run on the same input audio. We observe that streaming the LLM response as an input for the TTS component provides significant speedup to the total latency (Total (s) in Table \ref{tab:time_profile}) experienced between end of speech (as determined by VAD) and playing the first audio chunk generated

\subsection{TTS Component}

The main novelty of our work lies in the optimization of the TTS component. The results mentioned henceforth exhibit the improvements brought forth by our methodology.

First, we present results for which we utilize the same number of Mimi codebooks as the number of RVQ Iterations
Table~\ref{tab:gpu_cpu} summarizes the metrics for streaming TTS generation in GPU and CPU compute environments (with 32 RVQ Iterations) with the sentence "\textit{I am an AI, I am designed to assist and provide helpful responses to your queries. I am a machine learning model, trained on a vast amount of text data}" with no prior context being used for benchmarking and no LLM streaming. GPU compute shows a significant speedup compared to CPU and demonstrates the capability for real-time generation with RTF $<1$ and stable streaming output.\\
\newline
Table~\ref{tab:codebook_cpu} and \ref{tab:codebook_gpu} summarize the impact of number of RVQ Iterations in both CPU and GPU compute environments. Rows represent various relevant metrics and columns represent the number of RVQ Iterations.
There are common trends observed across both the environments and reflect that decreasing the number of RVQ Iterations improved compute times at the cost of audio quality.
\begin{itemize}
    \item Overall time to generate decreases as indicated by decrease in RTF values with decrease in RVQ Iterations. CPU environment achieves a usable RTF $<1$ only with 16 Iterations against the 32 intended 
    \item The latency to first chunk decreases with a decrease in RVQ Iterations, indicating improved responsiveness and user experience in field implementations
    \item As expected, the quality of audio decreases with a decrease in RVQ Iterations. The Signal to Noise Ratio (SNR) is computed using WADA-SNR \cite{wada-snr}
    \item Behavior of the TTS is quite similar for 20 and 24 RVQ Iterations, indicating a possible affinity to number of iterations being a multiple of 8 \\
\end{itemize}
The impact of number of RVQ iterations is much more pronounced in the GPU compute environment due to the fact that RVQ iterations involve a lot of sequential computations instructed by standard Python code and PyTorch operations resulting in several handoffs between GPU and CPU computation. This adds to the overhead and the CPU computation bottlenecks the GPU.
Next, we present the results of using 32 codebooks with Mimi decoder. 
Table \ref{tab:mimi} presents the quality of the voice outputs in terms of SNR in dB computed using WADA-SNR on GPU compute.
While the latency and RTF numbers remain almost similar to reduced Mimi codebooks supported by the fact that Audio decoding is not one of the bottlenecks of the TTS component, the deciding factor between the possible configurations turns out to be the audio quality.

\begin{figure}[htbp]
    \includegraphics[width=0.5\textwidth]{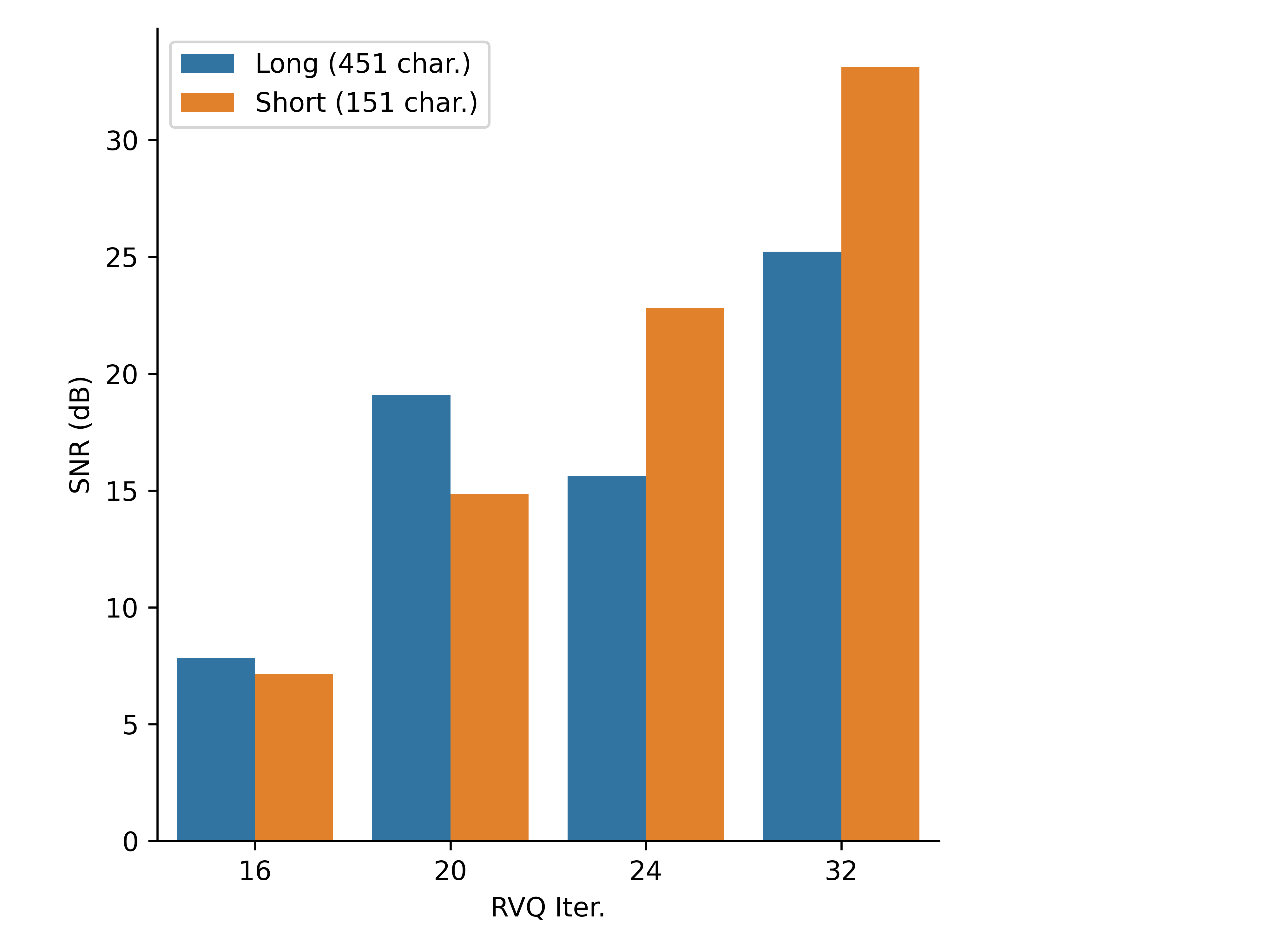}
    \caption{Effect of Audio Length on Quality (GPU): For lesser RVQ Iterations long audio clips have higher SNR values and hence better quality however as the number of RVQ Iterations approach the desired value the shorter audio has the higher SNR value indicating lesser noise}
    \label{fig:len_gpu}
\end{figure}

\begin{figure}[htbp]
    \includegraphics[width=0.5\textwidth]{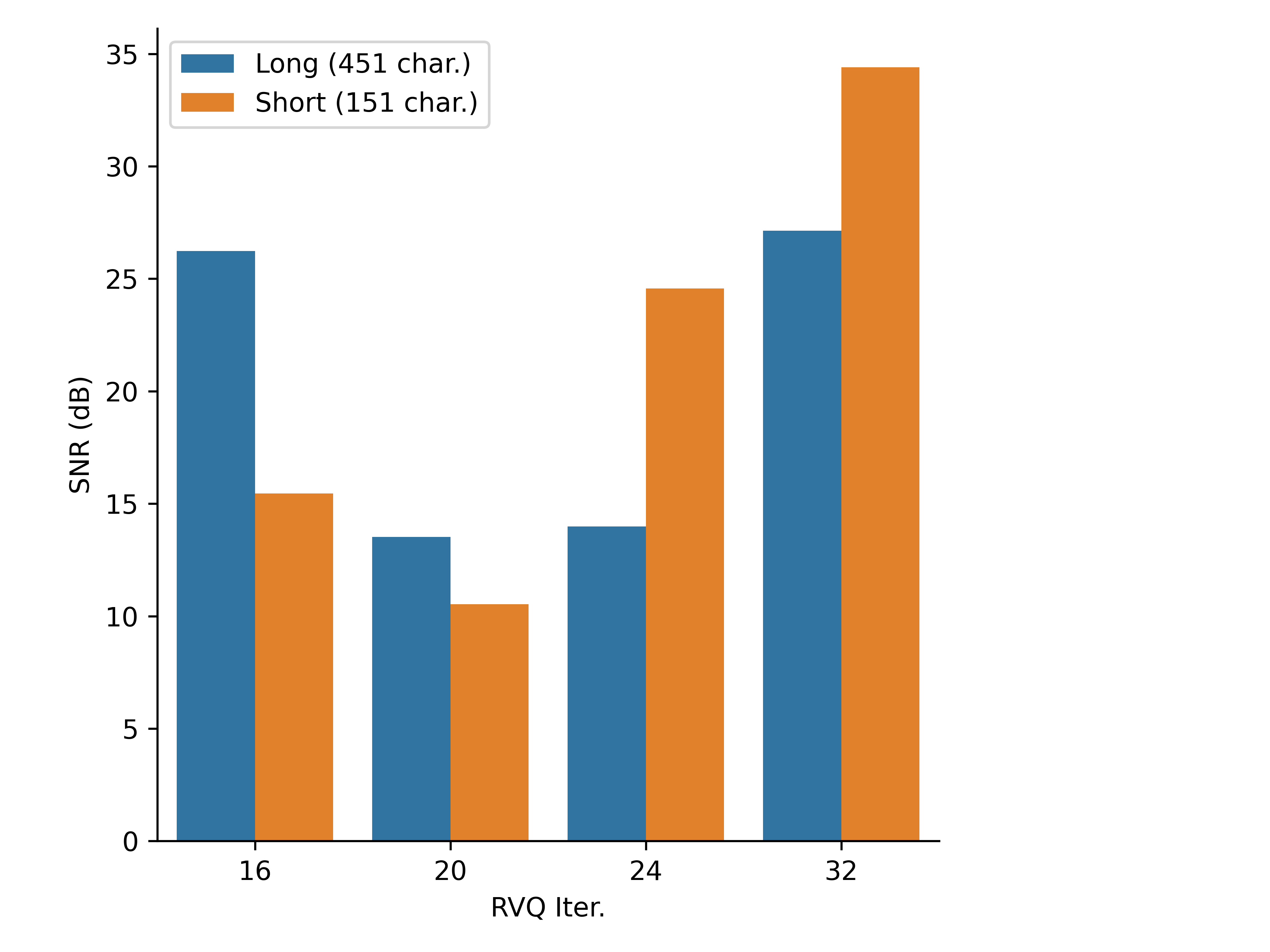}
    \caption{Effect of Audio Length on Quality (CPU): The effect is much more pronounced than the GPU environment with stability being observed at 16 and 32 RVQ iterations for longer audio clips}
    \label{fig:len_cpu}
\end{figure}

One noticeable effect on voice output is that it is elongated in length due to addition of several natural pauses in speech, which deters the use of 32 codebooks with reduced RVQ iterations for use-cases where the speech should sound natural. Observing several audio clips generated, there is high variance in SNR values going as low as -2.2341 (for 16 RVQ iterations with Concat Padding) the cause of which is unclear.
Further, we study the effect of length of audio generated on audio quality. Figure \ref{fig:len_gpu} and \ref{fig:len_cpu} depicts the effect of number of input text characters (451 vs 151) on audio quality.
As expected, the average size of a chunk and hence the latency to first chunk scale with the increase in length of audio generated, however the Real Time Factor (RTF) for generation stays the same demonstrating consistent efficiency.

\subsection{End-to-End Pipeline}
We utilize Silero-vad for robust speech detection and passing only speech sections of audio to the ASR module for efficient processing. Post speech detection, a 1.5 second window of silence was observed to be optimal for considering end of speech. Against a suite of LLMs, gpt-4o-mini provided a good balance of latency and token throughput. Our V-2-V architecture hosted in GPU enviroment shows a stable streaming voice output with Average Inter-Chunk Latency being close to two-thirds of the Average Chunk Length ensuring the stream queue is not empty post first chunk generation. However, V-2-V architecture hosted in CPU environment has noticeable gaps between streaming some chunks even when using 16 RVQ iterations.
\begin{table}[htbp]
\caption{Inference Metrics}
\centering
\begin{tabular}{|l|c|c|}
\hline
\textbf{Metric} & \textbf{GPU} & \textbf{CPU} \\
\hline
First Chunk Latency (ms) & 1021.2 & 2121.9  \\
Real-Time Factor (RTF) & 0.693x & 1.138x \\
Average Chunk Size (ms) & 1550.0 & 1544.0  \\
Average Inter-Chunk Latency (ms) & 1008.9 & 1660.4  \\
Min Inter-Chunk Latency (ms) & 821.5 & 1340.1  \\
Max Inter-Chunk Latency (ms) & 1319.7 & 1767.6  \\
Chunks per Second & 0.93 & 0.57 \\
\hline
\end{tabular}
\label{tab:gpu_cpu}
\end{table}

\begin{table}[htbp]
\caption{Optimization of RVQ Iterations on CPU}
\centering
\begin{tabular}{|l|c|c|c|c|}
\hline
\textbf{Metric (RVQ Iterations)} & \textbf{16} & \textbf{20} & \textbf{24} & \textbf{32} \\
\hline
First Chunk Latency (ms) & 1748.6 & 1855.0  & 2172.3  & 2662.5  \\
Real-Time Factor (RTF) & 0.934x & 1.143x & 1.236x & 1.489x \\
Number of Chunks & 5 & 7 & 7 & 7 \\
Avg. Chunk Size & 1504.0 & 1565.7  & 1394.3  & 1531.4  \\
Avg. Inter-Chunk Latency (ms) & 1191.3 & 1693.7 & 1563.2 & 2131.6 \\
Min Inter-Chunk Latency (ms) & 976.8 & 1542.8  & 367.8  & 1816.6  \\
Max Inter-Chunk Latency (ms) & 1306.1 & 2035.0  & 1879.2  & 2280.0  \\
Chunks per Second & 0.71 & 0.56 & 0.58 & 0.44 \\
SNR (dB) & 15.459  & 10.525 & 24.569  & 34.404 \\

\hline
\end{tabular}
\label{tab:codebook_cpu}
\end{table}

\begin{table}[htbp]
\caption{Optimization of RVQ Iterations on GPU}
\centering
\begin{tabular}{|l|c|c|c|c|}
\hline
\textbf{Metric (RVQ Iterations)} & \textbf{16} & \textbf{20} & \textbf{24} & \textbf{32} \\
\hline
First Chunk Latency (ms) & 640.9  & 1105.4  & 1172.3  & 1381.9  \\
Real-Time Factor (RTF) & 0.480x & 0.571x & 0.574x & 0.785x \\
Number of Chunks & 6 & 7 & 13 & 6 \\
Avg. Chunk Size (ms) & 1600.0  & 1565.7  & 1513.8  & 1373.3  \\
Avg. Inter-Chunk Latency (ms) & 685.1  & 773.0  & 800.4  & 915.2  \\
Min Inter-Chunk Latency (ms) & 597.6  & 701.3  & 47.8  & 251.0  \\
Max Inter-Chunk Latency (ms) & 949.0  & 1065.4  & 1344.7  & 1333.2  \\
Chunks per Second & 1.30 & 1.12 & 1.15 & 0.93 \\
SNR (dB) & 7.158 & 14.844 & 22.817 & 33.115 \\
\hline
\end{tabular}
\label{tab:codebook_gpu}
\end{table}

\begin{table}[htbp]
\caption{Audio Quality in SNR vs Audio Decoder Configurations}
\centering
\begin{tabular}{|l|c|c|c|c|}
\hline
\textbf{Configuration (RVQ iterations)} & \textbf{16} & \textbf{20} & \textbf{24} \\
\hline
32 Mimi Codebooks + Mean Pad & 7.3653 & 10.4672 & 26.49277  \\
32 Mimi Codebooks + Concat Pad & 8.5714 & 14.558 & 24.6396  \\ 
n Mimi Codebooks & 7.1582 & 14.8440 & 22.8174 \\
\hline
\end{tabular}
\label{tab:mimi}
\end{table}

\section{Conclusion}
Our work presents a voice-to-voice model focusing on low latency and delivering contextually aware expressive responses imperative in certain domains such as Customer Support. The focus of our experiments is around optimizing the CSM-1B TTS model, which helps effectively decreasing the latency to half through experimentation with the number of RVQ iterations presented in the original work. This has an impact on quality of speech output which may be insignificant for telephone based agent conversations where expectation on speech quality are not very high owing to the nature of the medium.  The important insights from our experiments on V-2-V architecture are as follows:

\begin{itemize}
    \item One-shot LLM response generation time remains a bottleneck for real-time conversation
    \item Streaming the LLM response substantially decreases the total latency by enabling the TTS component to generate audio chunks parallel to the LLM response generation
    \item Time taken by the ASR component may be absorbed into the silence window of the VAD component by enabling chunk based generation to process the audio chunks as they are recorded to essentially result in near zero effective latency contributed by the ASR component
    \item First audio chunk latency from TTS component can be optimized at the cost of audio quality
    \item First audio chunk latency scales proportionately to size of the audio chunk while maintaining a chunk size agnostic RTF, hence first chunk latency may be optimized by taking streaming input from the LLM component
\end{itemize}

\subsection*{Future Work}
The Residual Vector Quantization process in its current state is sequential Python code which may be optimized further by generating CUDA Kernels for the process so that it may natively run on GPU improving the latency and the user experience. Further experimentation may be conducted to replace the Llama 3.2 Backbone and Decoder used natively by the CSM model with more efficient models to further reduce the TTS latency.

The ASR generation may also be streamed so that the component may start processing the audio input as soon as it gets a positive from VAD instead of waiting for the entire input audio to be first collected by VAD and then sent to the ASR component. Implementation of ASR streaming would ensure that the ASR processing time is enclosed within the window of silence we set for VAD to detect end of user speech input hence providing 0 latency for the ASR component

\section{Acknowledgement}
The authors acknowledge  Amitabh Misra  and Yoginkumar Patel for their continuous encouragement to drive innovation through research in AI. Special thanks to other Sprinklr AI team members Saurabh Singh and Amit Kesari for their  support in setting up ML infra of this work.



\bibliographystyle{plain}
\bibliography{references}

\end{document}